\documentclass{article}%
\usepackage{amsmath}
\usepackage{amsfonts}
\usepackage{amssymb}
\usepackage{graphicx}%
\providecommand{\U}[1]{\protect\rule{.1in}{.1in}}
\begin{document}

\title{\vspace{-0.5in}\hfill{\normalsize UMTG - 5}\bigskip\\Euler Incognito}
\author{Thomas L Curtright$^{\text{\S ,\#}}$ and David B Fairlie$^{\text{\dag,\#}}%
${\large \medskip}\\$^{\text{\S }}${\small Department of Physics, University of Miami, Coral
Gables, Florida 33124, USA}\\$^{\text{\dag}}${\small Department of Mathematical Sciences, Durham
University, Durham, DH1 3LE, UK}}
\date{}
\maketitle

\begin{abstract}
The nonlinear flow equations discussed recently by Bender and Feinberg are all
reduced to the well-known Euler equation after change of variables.

\bigskip

\bigskip

\end{abstract}

Consider for $u\left(  x,t\right)  $ the class of nonlinear PDEs introduced
and discussed by Bender and Feinberg \cite{BenderFeinberg}.
\begin{equation}
u_{t}=\left(  u_{x}\right)  ^{k}u \label{Euler^k}%
\end{equation}
where $k$ is a parameter. $\ $(Including phases for the variables requires
only minor changes in the discussion to follow.)\ \ We are actually interested
only in $0\neq k\neq1$, since $k=0$ and $k=1$ are easily understood and
well-known. \ So, upon changing dependent variable to
\begin{equation}
v=\frac{k-1}{k}~u^{k/\left(  k-1\right)  } \label{1stVariableChange}%
\end{equation}
equation (\ref{Euler^k}) becomes%
\begin{equation}
v_{t}=\left(  v_{x}\right)  ^{k}\ . \label{Euler^kAgain}%
\end{equation}
Differentiating this with respect to $x$ and making a further change of
variable to
\begin{equation}
w=k\left(  v_{x}\right)  ^{k-1}=k\left(  u_{x}\right)  ^{k-1}u
\label{2ndVariableChange}%
\end{equation}
we find the familiar Euler-Monge equation in canonical form
\begin{equation}
w_{t}=w~w_{x}\ . \label{Euler}%
\end{equation}
Thus for any $k\neq0$ and $k\neq1$ the original equation (\ref{Euler^k}) is
reduced to Euler's through a change of dependent variable, although for
technical reasons that are more or less obvious from the explicit
constructions, it is often useful to assume $k>1$. \ (For $k=1$ of course,
(\ref{Euler^k}) is already the Euler-Monge equation \emph{without} any change
of variable.)

As is well-known (cf. \cite{BenderFeinberg}\ or \cite{CurtrightFairlie}\ for
references) the general solution for $w$ is given implicitly by%
\begin{equation}
w=F\left(  x+wt\right)
\end{equation}
where $F$ is an arbitrary differentiable function. \ By using the previous
changes of variables and integrating once with respect to $x$, solutions for
$v$, and hence $u$, follow from those for $w$. \ For example, if $F$ is
linear, $w=\frac{x-x_{0}}{t_{0}-t}$, $v=\frac{k-1}{k}\frac{\left(
x-x_{0}\right)  ^{k/\left(  k-1\right)  }}{\left[  k\left(  t_{0}-t\right)
\right]  ^{1/\left(  k-1\right)  }}$, and $u=\frac{x-x_{0}}{\left[  k\left(
t_{0}-t\right)  \right]  ^{1/k}}$.

\vfill

\noindent\underline
{$\ \ \ \ \ \ \ \ \ \ \ \ \ \ \ \ \ \ \ \ \ \ \ \ \ \ \ \ \ \ \ \ \ \ \ \ \ \ \ \ \ \ \ \ \ \ \ \ \ \ \ \ \ \ \ \ \ \ \ \ \ \ \ \ \ \ \ \ \ \ \ \ \ $%
}

$^{\text{\#}}${\footnotesize curtright@physics.miami.edu \&
david.fairlie@durham.ac.uk}

\newpage

\ \bigskip

\ \bigskip

Moreover, (\ref{Euler^k}) leads to two infinite families of local conserved
currents whose time and space components are powers of $u$ and $u_{x}$, but
not higher derivatives. \ The first family is quickly seen to be%
\begin{equation}
\left(  J_{0}^{\left(  n\right)  }~,~J_{1}^{\left(  n\right)  }\right)
=\left(  u^{n}u_{x}~,~u^{n+1}\left(  u_{x}\right)  ^{k}\right)
\end{equation}
for any $n$, not necessarily integer. \ Obviously, all these currents have
simple topological charges. \ On the solution set of (\ref{Euler^k}), $\left(
n+1\right)  J_{\mu}^{\left(  n\right)  }=\varepsilon_{\mu\nu}\partial^{\nu
}u^{n+1}$, and $\partial^{\mu}J_{\mu}^{\left(  n\right)  }=0$ immediately
follows. \ The second family of currents may be obtained from the known
(non-topological) conserved currents for $w$, namely $\left(  \left(
n+1\right)  w^{n}~,~nw^{n+1}\right)  $, just by changing variables.
\ Thus\footnote{In terms of $u$, as $k\rightarrow1$ we note that $\left(
n+1\right)  \left(  n+2\right)  J_{\mu}^{\left(  n\right)  }\rightarrow
\partial_{x}K_{\mu}^{\left(  n+1\right)  }\rightarrow\left(  \left(
n+2\right)  \left(  u^{n+1}\right)  _{x}~,~\left(  n+1\right)  \left(
u^{n+2}\right)  _{x}\right)  $.}%
\begin{align}
\left(  K_{0}^{\left(  n\right)  }~,~K_{1}^{\left(  n\right)  }\right)   &
=\left(  \left(  n+1\right)  \left(  v_{x}\right)  ^{n\left(  k-1\right)
}~,~nk\left(  v_{x}\right)  ^{\left(  n+1\right)  \left(  k-1\right)  }\right)
\nonumber\\
&  =\left(  \left(  n+1\right)  \left(  u_{x}\right)  ^{n\left(  k-1\right)
}u^{n}~,~nk\left(  u_{x}\right)  ^{\left(  n+1\right)  \left(  k-1\right)
}u^{n+1}\right)  \ .
\end{align}
On the solution set of (\ref{Euler^k}), or equivalently (\ref{Euler^kAgain}),
it is straightforward to show $\partial^{\mu}K_{\mu}^{\left(  n\right)  }=0$.

Finally, the linearization of (\ref{Euler}) as given in
\cite{CurtrightFairlie} can be used to relate the spatial derivative of
(\ref{Euler^kAgain}), or equivalently of (\ref{Euler^k}), to a linear
equation. \ Define%
\begin{equation}
\psi\left(  a,x,t\right)  \equiv\frac{1}{a}\left(  \exp\left(  ak\left(
v_{x}\right)  ^{k-1}\right)  -1\right)  \ .\label{encoding}%
\end{equation}
Technically, it is useful to assume $k>1$ here, especially for slowly varying
$v$. \ It follows that%
\begin{equation}
\left(  \frac{\partial}{\partial t}-\frac{\partial^{2}}{\partial a\partial
x}\right)  \psi=\left(  v_{t}-\left(  v_{x}\right)  ^{k}\right)  _{x}\times
k\left(  k-1\right)  \left(  v_{x}\right)  ^{k-2}\exp\left(  ak\left(
v_{x}\right)  ^{k-1}\right)  \ .\label{linear}%
\end{equation}
Thus $\left(  \frac{\partial}{\partial t}-\frac{\partial^{2}}{\partial
a\partial x}\right)  \psi=0$ iff $\left(  v_{t}-\left(  v_{x}\right)
^{k}\right)  _{x}=0$. \ (If the second factor on the RHS of (\ref{linear}%
)\ were to vanish, for both positive and negative $a$, this would require
$k>2$, $v_{x}=0$, and hence also $\left(  v_{t}-\left(  v_{x}\right)
^{k}\right)  _{x}=0$.) \ Encoding initial data for the nonlinear system in the
form (\ref{encoding}) therefore allows the data to be evolved linearly.
\ Given a well-behaved solution to the linear equation for $\psi$, we may then
extract the nonlinear data at other times just by taking the limit
$v_{x}\left(  x,t\right)  =\left(  \frac{1}{k}\lim\limits_{a\rightarrow0}%
\psi\left(  a,x,t\right)  \right)  ^{1/\left(  k-1\right)  }$. \ Integrating
with respect to $x$ modulo a function of only the time variable yields $v$,
hence $u$. \ \bigskip

\noindent\textbf{Acknowledgment \ }This material is based upon work supported
by the National Science Foundation under Grant No. 0555603.

\end{document}